\documentclass{pasa}%

\title[The Parsec-Scale Morphology of Southern GPS sources]{The Parsec-Scale Morphology of Southern GPS Sources}
\author[P.G. Edwards and S.J. Tingay]{P.G. Edwards$^1$ \and S.J. Tingay$^{2}$ \\
\affil{$^1$Australia Telescope National Facility, CSIRO Astronomy and Space Science, Epping, NSW, Australia}%
\affil{$^2$International Centre for Radio Astronomy Research, Curtin University, Bentley, WA, Australia}}%
\jid{PASA}
\doi{10.1017/pas.\the\year.xxx}
\jyear{\the\year}

\begin{document}%
\begin{abstract}
Multi-frequency, multi-epoch ATCA observations of a  sample of AGN resulted
in the identification of 9 new candidate Giga-hertz Peaked Spectrum
(GPS) sources.  Here we present Long Baseline Array observations at
4.8\,GHz of the four candidates with no previously published VLBI
image, and consider these together with previously published VLBI
images
of the other five sources. We find core-jet or compact double
morphologies dominate, with further observations required to
distinguish between these two possibilities for some sources.  One of the nine
candidates, PKS 1831$-$711, displays appreciable variability,
suggesting its GPS spectrum is more ephemeral in nature.  We focus in
particular on the apparent relationship between a narrow spectral
width and ``compact double'' parsec-scale morphology, finding further
examples, but also exceptions to this trend.  An examination of the
VLBI morphologies high-redshift ($z>$3) sub-class of GPS sources
suggests that core-jet morphologies predominate in this class.
\end{abstract}
\maketitle%
\section{INTRODUCTION }
\label{sec:intro}

GHz-Peaked Spectrum (GPS) sources are defined by a spectral turnover
at GHz frequencies
\cite{ode98}.
This class has been reported to represent at
least 10\% of the bright radio source population \cite{ode98} although
the more recent study by Mingaliev et al.\ \shortcite{min13} find that only
1$\sim$2\% of a sample of $\sim$5000 sources with $S_{\rm 5GHz}>$200\,mJy 
met their criteria for classification as a GPS source.  The catalog of
Labiano et al.\ \shortcite{lab07} contains 172 GPS sources, of which
156 have a flux density (generally, though not exclusively, at 5 GHz)
above 200\,mJy.

The class of GPS sources contains, amongst others, three distinct
sub-classes:
(i) radio galaxies, at low redshift,
(ii) quasars at relatively high redshift,
(iii) flaring active galactic nuclei which 
temporarily have a peaked spectrum during the spectral
evolution of an outburst.
Mingaliev et al.\ \shortcite{min13} found a quarter of 
their sample consisted of blazars with spectra that may temporarily had a
convex shape while the object was in an active state
(see also Lister 2003; Kovalev 2005; Torniainen et al.\ 2008).
While all these meet the basic definition of a GPS source, we
consider ``temporary'' GPS sources to be a separate class from
``bona fide'' or persistent GPS sources, and focus our interest on these.

In addition to the spectral criterion,
all three sub-classes of
GPS sources above are characterised
by compact sub-kpc--scale radio structure.
Bona fide GPS sources are further characterised by 
low radio polarisation, and
low variability at radio wavelengths.
An observed anti-correlation
between turnover frequency and projected linear size
\cite{fan90,ode97} has led to a model in which GPS sources evolve into
Compact Steep Spectrum (CSS) sources, which are typically less than
20\,kpc in projected linear size, and possibly then into FR-I and/or
FR-II sources \cite{sne03,tied15}.

The identification of the GPS source PKS 2000$-$330 with a $z=$3.77
quasar by Peterson et al.\ \shortcite{pet82} --- at the time most
distant source known --- led O'Dea \shortcite{ode90} to consider the
prevalence of GPS quasars at high redshift.  Although the population
of known $z>$3 sources at the time was not large, O'Dea found that
about half of the known $z>$3 quasars were GPS sources and
additionally that about half of the then known GPS quasars lay in this
redshift range.  More recently, Coppejans et al.\ \shortcite{cop15}
identified a sample of 33 megahertz peaked-spectrum (MPS) sources and
determined redshifts for 24, with an average redshift of 1.3.  Five of
the sources had $z>$2, with four fainter sources for which redshifts
could not be found, thought to be at even higher redshifts, suggesting
that the MPS sources are also good candidates for high-redshift
sources.  In contrast, however, Mingaliev et al.\ \shortcite{min13}
reported that there was a deficit of objects at large redshifts with
peak frequencies below several GHz.

Early studies of the parsec-scale morphology of GPS source found that
many could be characterised as a ``compact double'', with two
components of comparable flux density separated by up to some tens of
milliarcseconds.  Stanghellini \shortcite{sta03} concluded that GPS
quasars tend to have core-jet or complex morphologies whereas GPS
galaxies were more likely to be compact symmetric objects (CSO).

Data from a multi-frequency, multi-epoch ATCA survey \cite{tin03a} was
used to search for new GPS sources \cite{tin03b,edw04}.  The survey,
which was made in conjunction with the VSOP Survey Program
\cite{hir00}, originally contained 212 sources, however 17 of these
were dropped after the first few epochs as it was clear they did not
meet the criteria for inclusion in the VSOP Survey.  The remaining 185
sources were observed at up to 16 epochs between 1996 and 2000 at 1.4,
2.5, 4.8 and 8.6\,GHz.  The multi-epoch observations allowed
determination of the variability index, $M$, 
defined as the r.m.s.\ variation from the mean flux density,
divided by the mean flux density. 
Examination of these data
resulted in the identification of 9 new Southern Hemisphere GPS
candidates \cite{tin03b,edw04}.  In this paper we combine new LBA
observations
at 4.8\,GHz
of the four southernmost GPS candidates with published
images of the other 5 sources to study the parsec-scale morphologies
and the sizes of the compact components in these candidate GPS
sources, and consider the implications for the sub-categories of GPS
sources listed above.
In particular, we examine the correlation between
narrow spectral width and compact double morphology suggested by
Edwards \& Tingay \shortcite{edw04}, and the morphology of
the high redshift ($z>$3) sub-class of GPS sources
for insights into the nature of these sources.

\section{SOURCE CHARACTERISTICS}

We list in Table~\ref{tab1} the source classification and redshift,
and the 4.8\,GHz mean flux density, $S_5$, mean fractional
polarization, $p_5$, and variability index, $M_5$, 
from Tingay et al.\ \shortcite{tin03a}.  The width is the
FWHM of the fitted spectrum in decades of frequency (see Edwards \&
Tingay 2004 for details).

The mean variability index at 5\,GHz for the 185 sources ranged
between 0.01 and 0.43, with a median value of 0.08
(see Tingay et al.\
  2003a for tabulated values and distributions).
The highest
observed linear polarisation was 8.19\% at 5\,GHz, with a median level
of 2.2\%.  Twenty-four sources (13\% of the total) had polarisation
levels below the reliably measurable level of 0.5\%.  The nine new GPS
candidates identified by Edwards \& Tingay \shortcite{edw04}
showed the expected low variability and low fractional
polarization characteristic of the class.

Here we describe LBA observations of the four 
southernmost sources:
PKS 1619$-$680 (J1624$-$6809),
MRC 1722$-$644 (J1726$-$6427), 
PKS 1831$-$711 (J1837$-$7108), and 
PKS 2146$-$783 (J2152$-$7807). 
The other five candidates from Edwards \& Tingay \shortcite{edw04}
have declinations north of $-40^\circ$ and the results of
Very Long Baseline Array (VLBA) observations of
these sources have been reported previously.
Four of the five were observed at 5\,GHz 
as part of the VSOP Pre-Launch Survey 
(VLBApls: Fomalont et al.\ 2000), a companion program to the ATCA 
program to derive the final source list for the VSOP AGN Survey
\cite{hir00}, and the fifth source was studied by Tingay \& Edwards \shortcite{tied15}.

\begin{table*}
\caption[]{GPS candidates from Edwards \& Tingay \shortcite{edw04}.
Flux density at 4.8 GHz, $S_5$, variability index at 4.8 GHz, $M_5$, fractional polarisation
at 4.8 GHz, $p_5$, and peak frequency $\nu_{pk}$, are reproduced from Edwards \& Tingay \shortcite{edw04}.
LAS is the largest angular size inferred from the VLBI images considered in this paper.
See text for details.}\label{tab1}
\begin{tabular*}{\textwidth}{@{}l\x l\x c\x c\x c\x c\x c\x c\x r\x r\x r@{}}
\hline
\noalign{\smallskip}
Source         & J2000        &  $z$  & $\nu_{pk}$ & width & $M_5$ &  $p_5$ &  $S_5$ &  $S_{VLBI}$ & LAS    & LAS  \\ 
name           &              &       & (GHz)      &       & (\%)  &        &  (Jy)  &  (Jy)       & (mas)  & pc    \\
\noalign{\smallskip}
\hline
\hline
\noalign{\smallskip}
PKS 0150$-$334 & J0153$-$3310 & 0.610 &        1.5 &   1.6 &  0.04 &   1.35 &   0.88 &   0.86 &  1.0 &   6.8 \\ 
PKS 0434$-$188 & J0437$-$1844 & 2.702 &        4.5 &   1.3 &  0.05 &   0.80 &   0.95 &   1.08 &  1.1 &   9.1 \\ 
PKS 0642$-$349 & J0644$-$3459 & 2.165 &        3.3 &   1.4 &  0.09 &   2.03 &   0.85 &   0.84 &  3.4 &  28.7 \\ 
PKS 1619$-$680 & J1624$-$6809 & 1.360 &        3.1 &   1.4 &  0.04 & $<$0.5 &   1.69 &   1.56 &  4.0 &  34.2 \\ 
MRC 1656$-$075 & J1658$-$0739 & 3.742 &        4.8 &   1.2 &  0.03 & $<$0.5 &   1.32 &   1.36 &  7.1 &  51.7 \\ 
MRC 1722$-$644 & J1726$-$6427 &  ...  &        1.1 &   1.0 &  0.02 & $<$0.5 &   1.26 &   1.36 & 47.0 &   ... \\ 
PKS 1831$-$711 & J1837$-$7108 & 1.356 &        8.2 &   2.0 &  0.06 &   1.77 &   2.39 &   3.40 & 13.4 & 114.5 \\ 
PKS 2146$-$783 & J2152$-$7807 & 3.997 &        4.3 &   1.3 &  0.03 & $<$0.5 &   1.15 &   1.39 & 0.4  &   2.6 \\ 
PKS 2254$-$367 & J2257$-$3627 & 0.006 &        2.7 &   1.8 &  0.04 & $<$0.5 &   1.28 &   1.15 & 83.0 &  10.3 \\ 
\noalign{\smallskip}
\hline
\hline
\end{tabular*}
\end{table*}

\begin{figure}
\begin{center}
\includegraphics[width=\columnwidth,angle=-90]{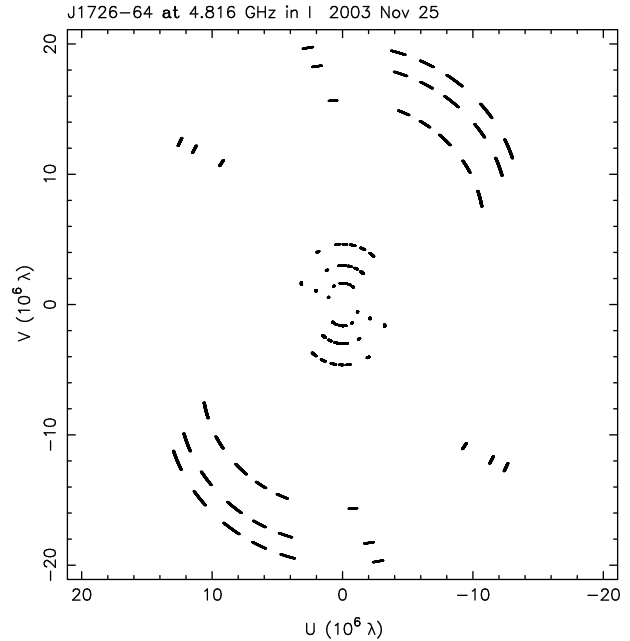}
\caption{The ($u,v$) coverage for MRC 1722$-$644 (J1726$-$6427) with Parkes, ATCA, Mopra and Hobart.
The scans were typically 15min in duration.}\label{Fig1}
\end{center}
\end{figure}

\section{LBA OBSERVATIONS}

The southernmost four GPS candidates were observed with the Australian
Long Baseline Array (Edwards \& Phillips 2015) over an 18 hour period
starting on 2003 November 26 at 20:00 UT (proposal code v170). The
observing schedule cycled around the four sources such that each
source was observed for approximately one quarter of the time but over
the full time range of the observation.  In addition to the target
sources, PKS 1921$-$293 (approximately 9.4 Jy at this time for our
observing frequency) and PKS 1610$-$771 (approximately 3.7 Jy at this
time for our observing frequency) were observed for short periods as
fringe-finders and to assist in establishing the flux density
calibration for the observations.
These flux densities were obtained
by interpolating between observations in that ATCA Calibrator
Database\footnote{http://www.narrabri.atnf.csiro.au/calibrators/}.

The array of telescopes consisted of
the Australia Telescope Compact Array (ATCA) (5\,x\,22\,m), Mopra
(22\,m), Parkes (64\,m), Hobart (26\,m), and Ceduna (30\,m).  The S2
recording system was used to record a 16\,MHz band in left and right
circular polarisations, centred at 4.816\,GHz.  The recorded data were
correlated at the Australia Telescope National Facility S2 correlator
\cite{wil96}.
Due to instrumental issues at Ceduna, the data from this telescope was
not usable and is not considered further.

The correlated data were imported into AIPS\footnote{AIPS is produced
  and maintained by the National Radio Astronomy Observatory, a
  facility of the National Science Foundation operated under
  cooperative agreement by Associated Universities, Inc.}.  As all the
target sources are bright, phase reference calibrators were not
included in the observing schedule.  Fringe-fitting proceeded on the
fringe-finders and targets, finding good delay and rate solutions for
each antenna across the full time range and for both polarisations.
The fringe-fit solutions were applied to the data.  System temperature
data and gain information from each antenna were imported into AIPS
and applied to the correlated data to convert the amplitudes from
correlation coefficients into jansky.  We examined the calibrated data for
PKS 1921$-$293 and estimated corrections to the antenna gains by
comparison to the cataloged flux density for PKS 1921$-$293 at 4.8\,GHz
near the observing epoch.  These corrections were then applied to all
target sources, via a second round of amplitude calibration in AIPS.

The data for all target sources were then exported as UVFITS files and
further examined using Difmap \cite{she97}.  In Difmap, the data for
each target was flagged for bad visibilities, averaged over 10 second
periods (the correlation integration time was 2 seconds), and averaged
the two circular polarisations into Stokes I.  As an example, the
($u,v$) coverage for
MRC 1722$-$644 (J1726$-$6427)
is shown in
Figure~\ref{Fig1}. Eaach target was imaged (images sizes of
128$\times$128 pixels of 3 mas pixel size) and calibrated using
standard clean and self-calibration tasks.  Deconvolution was guided
by clean windows, set on the peak residuals in the field, and cleaning
proceeded until the peak negative and peak positive residuals had
comparable magnitudes.  Both phase and amplitude self-calibration was
employed, with overall amplitude self-calibration corrections to the
flux densities of order 10\% for all antennas and all targets.
We
adopt 10\% as the error on the flux density estimates
guided by previous experience (e.g., Tingay et al.\ 1998).
The angular
resolution achieved in the images ranged between 6 and 23\,mas
(Figures~2 to 5).  The typical image r.m.s.\ achieved was approximately
10\,mJy, producing images with dynamic ranges of approximately 50 to
100.

In order to extract quantitative information on the target source
morphologies, at the end of the imaging and self-calibration process,
we replaced the point source clean component models 
with elliptical Gaussian components, between one and three for each
target source.
The Difmap task modelfit was then used to fit the
models to the data for each target.  The results of the model fitting
are given in Table~\ref{tab2}.

\begin{table*}
\caption{Parameters of the target source structure models (elliptical Gaussian model components).  
$S$ is the total flux density at 4.8\,GHz of the model component in Jy.  
$r$ is the angular distance of the model component from the image phase centre in milliarcseconds.  
$\theta$ is the position angle of the model component relative to the phase centre, in degrees east of north.  
$a$ is the full width at half maximum of the major axis of the model component in milliarcseconds.  
$b$ is the ratio of the minor axis to the major axis full widths at half maximum.  
$\phi$ is the position angle of the major axis of the model component, east of north.}\label{tab2}
\begin{center}
\begin{tabular*}{\textwidth}{@{}l\x l\x r\x r\x r\x r\x r\x r@{}}
\hline \hline
Source         & J2000       &   $S$   & $r$    & $\theta$  & $a$    &  $b$  &  $\phi$ \\ 
               &             &   (Jy)  & (mas)  & (deg)     & (mas)  &       &   (deg) \\ \hline
PKS 1619$-$680 &J1624$-$6809 &    0.71 &    5.1 &        56 &    6.7 &  0.35 &      55 \\
               &             &    0.85 &    0.9 &        80 &    5.4 &  0.00 &      66 \\ 
MRC 1722$-$644 &J1726$-$6427 &    0.53 &    6.4 &      $-$2 &    6.1 &  0.54 &      46 \\
               &             &    0.36 &    2.5 &        90 &    5.7 &  0.25 &      82 \\
               &             &    0.47 &   40.8 &       141 &    5.1 &  0.26 &      49 \\ 
PKS 1831$-$711 &J1837$-$7108 &     2.3 &    0.1 &     $-$62 &    2.1 &  0.64 &   $-$18 \\
               &             &     1.1 &   13.3 &       142 &    7.7 &  0.00 &      65 \\ 
PKS 2146$-$783 &J2152$-$7807 &    1.39 &    0.0 &     $-$26 &   0.36 &   1.0 &  $-$169 \\ 
\hline \hline
\end{tabular*}
\end{center}
\end{table*}

\section{RESULTS}

In this section we present the results of our LBA observations and consider
published observations of the other sources and attempt to characterise
the parsec-scale morphology of the source.
We are guided by the work of 
Snellen et al.\ \shortcite{sne00}
who classified the radio morphologies as
Core-Jet sources (CJ),
Compact Symmetric Objects (CSO),
Compact Double sources (CD),
or Complex sources.
Compact Doubles may be considered a subset of Compact Symmetric Objects
(see, e.g., Peck \& Taylor 2000): a CD source has no central component,
whereas a CSO has a central component with extended structure on both sides.

\noindent{\bf PKS 0150$-$334:}
Imaged at 5\,GHz as part of the VLBApls observations in June 1996 and
resolved into two components of similar brightness separated by 1\,mas
\cite{fom00}.
While at face value this is suggestive of a compact
double, the spacing is somewhat less than would be typical for this
class (cf.\ Snellen et al.\ 2000) and a core-jet morphology cannot be ruled out.
This is borne out by the subsquent observation at 8.4\,GHz by
Ojha et al.\ \shortcite{ojh04} in November 2002, which revealed two
components separated by 3.3\,mas.  Acknowledging that a speed derived
from only two observations should be treated with caution, at a
redshift of 0.61, an apparent motion of 2.3\,mas in 6.4 years
corresponds to an apparent superluminal motion of 12.8\,$c$.  Further
observations are required to confirm this but, in any case, it is
apparent that the source has a core-jet morphology.

\noindent{\bf PKS 0434$-$188:}
Imaged as part of the VLBApls \cite{fom00} and resolved into two
components of similar brightness (0.62 and 0.45 Jy) separated by
1.1\,mas.  The source was observed in February 2002 as part of the VSOP
Survey Program with the resulting model containing two components
(0.62 and 0.35 Jy) separated by 1.3\,mas \cite{dod08}.  (We note an
error in the formatting of Table~3 of Dodson et al.\ which has
resulted in a third component being erroneously listed for this source
--- closer inspection indicates it is actually the second component of
J0743$-$6726.)  In the light of the previous example, further
observations are required to better characterise the source evolution,
both spectral and morphological, and distinguish between a compact double or
core-jet morphology.

\noindent{\bf PKS 0642$-$349:}
The VLBApls image \cite{fom00} is dominated by a compact core of
0.69\,Jy and a secondary component of 0.15\,Jy located 3.4\,mas to the
west, with a possible third component a further 10\,mas to west. This
has the characteristics of a core-jet morphology.

\begin{figure}
\begin{center}
\includegraphics[width=\columnwidth,angle=-90]{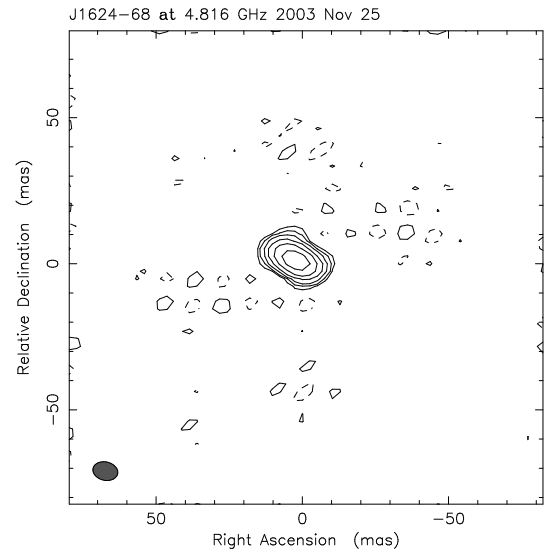}
\caption{LBA image of PKS 1619$-$680 (J1624$-$6809).
The image peak is 0.99 Jy/beam, with contours at
$-$2, 2, 4, 8, 16, 32 and 64 per cent of the peak.
The beam FWHM is 8.6\,mas $\times$ 6.3\,mas at a position angle of 77$^\circ$.}\label{Fig2}
\end{center}
\end{figure}

\noindent{\bf PKS 1619$-$680;}
Our LBA image (Figure~\ref{Fig2}) reveals two components of similar
flux density and with similar major axes
separated by 7\,mas. We note the modelfit to the data
presented in Table~\ref{tab2} has
a linear (formally zero-width)
second component: we do not
interpret this literally but rather take it to suggest that an
elogated Gaussian component would provide an acceptable fit, and that
a higher resolution observation may resolve this component into a
number of smaller components.  Higher resolution observations are
required to distinguish between a compact double morphology or a
core-jet morphology.

\noindent{\bf MRC 1656$-$075:}
The VLBApls image is dominated by a bright, resolved 1.3\,Jy
component, with a 0.05\,Jy secondary offset by 7.1\,mas \cite{fom00}.
The disparity in flux densities argues against classification as a
compact double --- Peck \& Taylor \shortcite{pec00}, for example,
adopted a requirement of a ratio less than 10:1 for a CSO --- and
suggests a core-jet morphology. Multi-frequency data to examine
the spectral characteristics of the components would be useful.

\begin{figure}
\begin{center}
\includegraphics[width=\columnwidth,angle=-90]{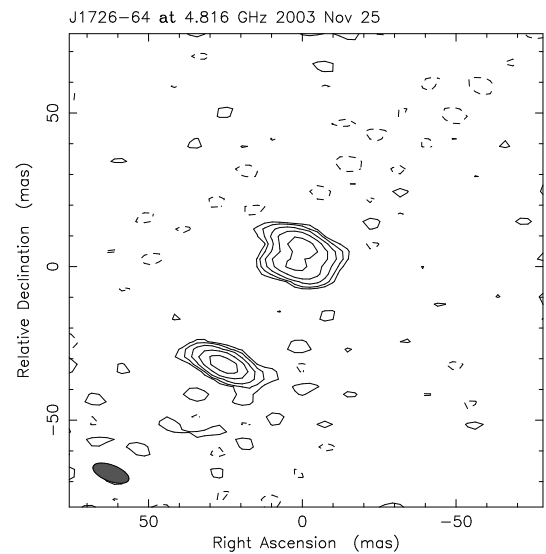}
\caption{LBA image of MRC 1722$-$644 (J1726$-$6427). 
The image peak is 0.42 Jy/beam, with contours at
$-$4, 4, 8, 16, 32 and 64 per cent of the peak.
The beam FWHM is 12.4\,mas $\times$ 5.3\,mas at a position angle of 69$^\circ$.}\label{Fig3}
\end{center}
\end{figure}

\noindent{\bf MRC 1722$-$644:}
Our LBA observations resolve this source into two components separated
by almost 40\,mas (Figure~\ref{Fig3}). The brighter of these is more
extended, and can be modelled by two overlapping components.  This
morphology is reminiscent of PKS 1934$-$638 \cite{tzi02,tzi10}, and we
categorise it as a compact double.  Dodson et al.\ \shortcite{dod08}
note that in a VSOP Survey observation there were no detections on
space baselines.

\begin{figure}
\begin{center}
\includegraphics[width=\columnwidth,angle=-90]{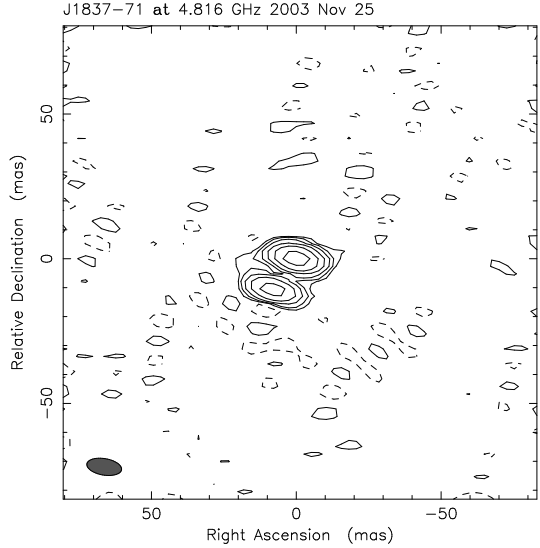}
\caption{LBA image of PKS 1831$-$711 (J1837$-$7108).
The image peak is 2.22 Jy/beam, with contours at
$-$2, 2, 4, 8, 16, 32 and 64 per cent of the peak.
The beam FWHM is 12.1\,mas $\times$ 5.7\,mas at a position angle of 80$^\circ$.}\label{Fig4}
\end{center}
\end{figure}

\noindent{\bf PKS 1831$-$711:}
Our LBA observations resolve this source into two components separated
by 13\,mas (Figure~\ref{Fig4}). The brighter component is about twice
as bright as the fainter, with the latter being more extended. The
best modelfit to the more extended component is a single linear
component which we consider in the same manner as for PKS 1619$-$680.
Based on the image alone, we would provisionally classify this as a
compact double, however as discussed in the next section,
consideration of additional information leads us to favour a core-jet
interpretation.  We note that a VSOP survey observation in May 1999
found a compact core, 0.2\,mas in size, and with evidence of some
extended structure \cite{dod08}, but suggest the limited ground array
in that observation (Hobart, Mopra, Hartebeesthoek) may have struggled
to resolve the structure revealed by our observation.

\begin{figure}
\begin{center}
\includegraphics[width=\columnwidth,angle=-90]{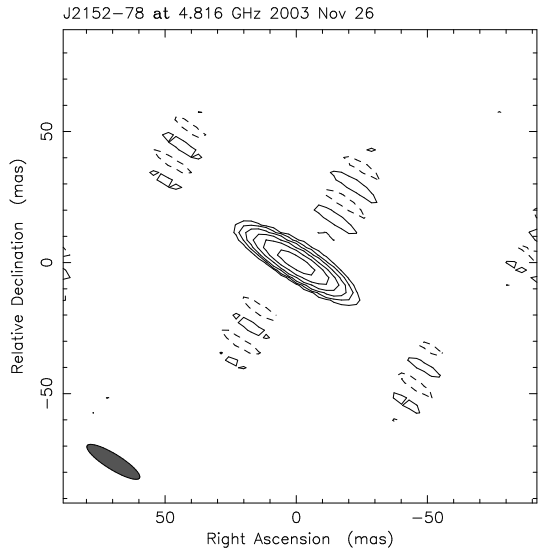}
\caption{LBA image of PKS 2146$-$783 (J2152$-$7807). 
The image peak is 1.39 Jy/beam, with contours at
$-$2, 2, 4, 8, 16, 32 and 64 per cent of the peak.
The beam FWHM is 23.5\,mas $\times$ 5.9\,mas at a position angle of 58$^\circ$.}\label{Fig5}
\end{center}
\end{figure}

\noindent{\bf PKS 2146$-$783:}
Our LBA observations yield a strongly core-dominated source, modelled
by a single circular Gaussian component (Figure~\ref{Fig5}).  Despite
this, Dodson et al.\ \shortcite{dod08} found no detections on space
baselines in a VSOP survey observation.  Based on our observations,
the source is unresolved. 

\noindent{\bf PKS 2254$-$367:}
This low-redshift GPS source was imaged at four frequencies with the
VLBA in November 2003, revealing a compact core and symmetric
structure on both sides, leading to classification as a compact
symmetric object \cite{tied15}.

\section{DISCUSSION}

\subsection{\bf Status of Southern GPS candidates}

In Table~\ref{tab1} we give the sum of the flux densities of the
components in the VLBI image and compare this to the mean flux density
in the ATCA monitoring.  We note these measurements are
non-contemporaneous (VLBApls in June 1996; ATCA monitoring 1996 to
2000; our LBA observations 2003) but as bona fide GPS sources display generally
low variability this is not as critical as might otherwise be the
case. It is apparent that the VLBI observations have recovered
$\gtrsim$90\% of the flux density in all cases. The ``missing'' flux
density may be due to low-level extended emission, or to source
variability.

The biggest discrepancy is for PKS 1831$-$711, where the VLBI flux
density signficantly exceeds the ATCA flux density.  Edwards \& Tingay
\shortcite{edw04} previously noted that the PKSCAT flux densities for
this source, e.g.\ 1.15\,Jy at 5.0\,GHz \cite{pkscat}), differ
significantly from more recent values, e.g., 2.29\,Jy \cite{wri94} and
2.39\,Jy \cite{tin03a}, suggesting significant long term variability.
The total flux density of the components in the LBA image, 3.4\,Jy, is
higher again, confirming this source is significantly more variable
than the majority of GPS sources. (Its ATCA variability index was the
second highest of the 9 candidates here.)  Edwards \& Tingay
\shortcite{edw04} also pointed out that the spectral width of
J1837$-$7108 was the largest of all 9 candidates.  This is further
evidence in favour of a core-jet interpretation of the morphology, and
suggests that the source belongs to the class of AGN which only
temporarily have a peaked spectrum during the spectral evolution of an
outburst.

Similarly, Edwards \& Tingay \shortcite{edw04} inferred moderate
long-term variability for J2152$-$7807 from differences between PKSCAT
values, e.g., 0.77\,Jy at 5\,GHz \cite{pkscat}, and more recent
determinations, 1.13\,Jy \cite{wri94} and 1.15\,Jy \cite{tin03a}.

Table~\ref{tab1} also contains the largest angular size (LAS) of the
sources based on the VLBI observations.  The conversion from angular
to projected linear size was made using the cosmology calculator of
Wright \shortcite{wri06} adopting $H_0$=69.6 km\,s$^{-1}$\,Mpc$^{-1}$,
$\Omega_M$=0.286 and $\Omega_{\rm vac}$=0.714.  The values of the LAS
confirm that the milli-arcsecond scale structure (which, as noted
above, makes up the overwhelming majority of the radio emission) is
confined within a kiloparsec in all cases, as expected for GPS
sources.

\subsection{Distribution of morphological types}

We have classified the parsec-scale morphologies as four likely
core-jet sources, one CSO, one compact double, one unresolved, and two
which may be compact doubles or core-jet sources.  Further
observations of these latter two, ideally at two or more frequencies
to determine the spectral characteristics of the components, would
enable their classifications to be determined.

Although our sample is small, it is interesting to compare it with the
results of Snellen et al.\ \shortcite{sne00}. For their sample of 47
faint GPS sources (with peak flux densities ranging between 30 and
900\,mJy), 3 were classified as CSO, 11 as CD, 7 as CJ, and 2 as
complex.  Of the remainder, 20 could not be classified as they were
too compact.  Snellen et al.\ \shortcite{sne00} noted, as we have,
that many of the sources classified as CD or CJ had only two
components, resulting in their classifications being tentative.  The
biggest difference between our results and those of Snellen et al.\ is
that we have far fewer unresolved sources, most likely as the sources
in our sample are brighter, with peak flux densities $\gtrsim$1\,Jy.

Stanghellini et al.\ \shortcite{sta97,sta99} considered the morphological
information available for a sample of 20 GPS sources and found that
GPS quasars have preferentially core-jet or complex morphologies and GPS
galaxies tend to be CSOs.  Only one of our candidates is classified as
a galaxy, and it is a CSO.  We find less evidence for complex
morphologies among the quasars, however our sample is small.

We note that PKS 0642$-$349, which has a clear core-jet morphology, has the
highest fractional polarisation (2\%) and highest variability index
(0.09) of the nine candidates. PKS 0150$-$334, which also has a
core-jet morphology, similarly has a detectable fractional polarisation.

\subsection{High redshift sources}

Edwards \& Tingay \shortcite{edw04} considered the redshift
distribution of the GPS sources and candidates and confirmed it
differed significantly from that of a larger flat-spectrum sample. The
trend has become even stronger for this sample of 9 candidates, as two
of the three sources without redshift determinations at the time have
subsequently been identified by Healey et al.\ \shortcite{hea08} with
$z>$3 quasars.  (The other, MRC 1722$-$644, still has no
redshift.)

Mingaliev et al.\ \shortcite{min13} report that there is a deficit of
objects at large redshifts with peak frequencies below several GHz.
Inspection of Table~\ref{tab1} appears to support this, as the $z>$2
sources all have peak frequencies greater than 3\,GHz, however
comparison with the larger sample of sources in Table~1 of Edwards \&
Tingay \shortcite{edw04} reveals this apparent trend is not as evident
when a larger sample (albeit smaller than that of Mingaliev et al.) is
considered.

\begin{table*}
\caption[]{GPS sources with redshifts, $z>$3, from O'Dea (1990), Labiano et al.\ (2007) and this work.
Morphologies are U (unresolved) or CJ (core-jet).
LAS is the approximate largest angular size (see text for details). }\label{tab3}
\begin{tabular*}{\textwidth}{@{}l\x l\x c\x c\x r\x l\x l\x  r@{}}
\hline
\noalign{\smallskip}
B1950      & J2000    & $z$ & Morphology & LAS   & Reference \\ 
           &          &     &            & (mas) &           \\ 
\noalign{\smallskip}
\hline
\hline
\noalign{\smallskip}
0201$+$113 & J0203$+$1134 & 3.639 & CJ   & 2   & Pushkarev \& Kovalev \shortcite{push12}     \\ 
0420$-$388 & J0422$-$3844 & 3.110 & ...  & ... &   ...                 \\ 
0636$+$680 & J0642$+$6758 & 3.180 & U    & ... & Britzen et al.\ \shortcite{bri07}, Pushkarev \& Kovalev \shortcite{push12}  \\ 
0642$+$449 & J0646$+$4451 & 3.396 & CJ   & 9   & Gurvits et al.\ \shortcite{gur92}        \\ 
1351$-$018 & J1354$-$0206 & 3.707 & CJ   & 13  & Frey et al.\ \shortcite{fre97}        \\ 
1354$-$174 & J1357$-$1744 & 3.147 & CJ   & 32  & Frey et al.\ \shortcite{fre97}        \\ 
1422$+$231 & J1424$+$2256 & 3.626 & CJ   & ... & Orienti et al.\ \shortcite{ori06}        \\ 
1442$+$101 & J1445$+$0958 & 3.535 & CJ   & 22  & Pushkarev \& Kovalev \shortcite{push12}       \\ 
1526$+$670 & J1526$+$6650 & 3.020 & CJ   & 2   & Britzen et al.\ \shortcite{bri07}  \\ 
1614$+$051 & J1616$+$0459 & 3.197 & CJ?  & 2   & Orienti et al.\ \shortcite{ori06}, Pushkarev \& Kovalev \shortcite{push12}  \\ 
1656$-$075 & J1658$-$0739 & 3.742 & CJ   & 7   &   this work           \\ 
1839$+$389 & J1840$+$3900 & 3.095 & CJ   & 2   & Britzen et al.\ \shortcite{bri07}     \\ 
2000$-$330 & J2003$-$3251 & 3.773 & CJ   & 2   & Ojha et al.\ \shortcite{ojh05}        \\ 
2126$-$158 & J2129$-$1538 & 3.270 & CJ   & 2   & Pushkarev \& Kovalev \shortcite{push12}   \\ 
2146$-$783 & J2152$-$7807 & 3.997 & U    & ... & this work           \\      
\noalign{\smallskip}
\hline
\hline
\end{tabular*}
\end{table*}

In Table~\ref{tab3} we compile a list of GPS sources and canididates
with $z>$3 from O'Dea \shortcite{ode90}, Labiano et
al.\ \shortcite{lab07} and this work, in order to examine the
parsec-scale morphologies of this class of extreme GPS sources.  The
age of universe spanned by these redshifts corresponds to 1.6 to 2.2
Gyr.  B0429$-$388 is the only source for which there is not a
published VLBI image (and we note that while the spectrum is curved it
has yet to be firmly established that it is in fact peaked).  We
tabulate the approximate Largest Angular Size of the parsec-scale
image, noting that the VLBI images span a range of sensitivities due
to both the epoch the observations were made (more recent observations
are generally made with wider bandwidths) and the observing mode (some
are snapshots, others are dedicated observations). We do not give a
LAS for B1422+231 as it is gravitational lensed and the lensed images
are subject to a range of amplifications in flux density and scale,
though none of the images is larger than several mas in extent
\cite{ori06}.

Only two of the sources are unresolved, and several extend over more
than 10\,mas. The resolved sources all appear to have a core-jet
morphology: Orienti et al.\ \shortcite{ori06} proposed B1614+051 as a
candidate CSO source, however comparison with the subsequent observation of Pushkarev
and Kovalev \shortcite{push12} lead us to tentatively favour a
core-jet morphology.  This predominance of core-jet morphologies is in
accord with previous reports that quasars are more likely to have
core-jet or complex structures \cite{sta03}.

We note that the sources in Table~\ref{tab3} are compiled from
several sources and are subject to their respective selection biases,
and reiterate that the VLBI morphologies are based on a heterogeneous
set of VLBI observations. Nevertheless, the predominance of CJ
morphologies, and absence of CD or CSO morphologies, is striking
and deserves a more systematic follow-up.

\subsection{The spectral widths of Compact Doubles}

Edwards \& Tingay \shortcite{edw04} observed that the spectrum of
MRC 1722$-$644 has a width comparable to the (narrow) spectra of the
compact double PKS 1934$-$638 \cite{tzi02,tzi10} and the compact
symmetric object 0108+388 \cite{ode91,mar01} which led them to predict
that MRC 1722$-$644 might also display a Compact Double morphology,
which is borne out by our LBA observations.

Further examples can be found in the sample of Dallacasa et
al.\ \shortcite{dal98}: 1518+047 and 1607+268 are Compact Doubles with
separations of 150 and 60\,mas respectively, and both have spectral
widths of $\sim$1.1 decades of frequency.  More recently, Callingham
et al.\ \shortcite{cal15} reported that the GPS source PKS B0008$-$421
has the smallest known spectral width of any GPS source, and the
steepest known spectral slope below the turnover, close to the
theoretical limit of synchrotron self-absorption.  The VLBI-scale
morphology has two components separated by $\sim$120\,mas.

Simplistically, such a correlation seems reasonable: if compact double
sources are comprised of two lobes, with the central component either
heavily absorbed (at these frequencies) or in a quiescent state, it
would be quite natural for the superposition of the spectra of the two
components to be narrower than the case where an additional, generally
flatter spectrum, core component also contributes to the total source
spectrum. However, inspection of other compact double sources
indicates that any such correlation is not tight: Taylor \& Peck
\shortcite{tay03} found PKS 2344$-$192 (J2347$-$1856) to be a compact
double with a separation of $\sim$32 mas, but the source has yet to be
established as a GPS source (while the spectrum is curved, it is not
clear it is peaked).  Similarly, Taylor et al.\ \shortcite{tay00}
found 1031+567 to be a compact double with a separation of $\sim$34
mas. While archival data indicates it has a peaked spectrum, the width
is $\sim$1.5 decades of frequency, appreciably broader than the
examples above.

Generally, however, sources with CD or CSO morphologies must either be
viewed close to side-on, or possess jets where doppler boosting does
not play a significant role, if we are indeed seeing two lobes of
similar brightness.  Strong absorption of the core suggests we are
viewing the source side-on, with the absoption in a disc or torus of
material surrounding the core.  In contrast, in core-jet sources ---
assuming jets are intriscially two-sided --- we are seeing the doppler
boosted jet oriented towards us and not seeing the doppler de-boosted
jet travelling in the opposite direction (although again, absorption
could play a role close to the core).  Variability is a common feature
of core-jet sources, amplified by Doppler boosting, and so one might
expect core-jet sources generally to be more variable, and therefore
for the spectra of core-jet sources to evolve away from having GPS
spectra more quickly than their CSO or CD counterparts.  Compact
Double sources might then be expected to be less (or more slowly)
variable, as is the case for 1031+567 \cite{fas01} and PKS 1934$-$638,
the primary flux density calibrator for the ATCA \cite{par16}. MRC
1722$-$644, which we classify as a CD, has the lowest variability
index of the nine candidates (see Table~\ref{tab1}).

\section{FUTURE WORK}

We have established that one of the candidate GPS sources examined
shows evidence for sufficient variability to exclude it as a bona fide
GPS source.  The other sources remain good candidates for inclusion in
GPS catalogues.  Multi-epoch, and multi-frequency, VLBI observations
in the future would enable the morphologies of the sources to
be established more firmly, particular in discriminating between
core-jet sources and compact double sources.

On-going montoring to establish the persistence (or otherwise) of the
peaked spectra is another approach, the utility of which has been
demonstrated for PKS 1718$-$649 by Tingay et al.\ \shortcite{tin15}.
In addition, results at lower frequencies from the MWA GLEAM survey
(Wayth et al.\ 2015; Hurley-Walker et al.\ in preparation) and the
Australian SKA Pathfinder \cite{jon07} will extend the frequency range
over which the spectra can be determined and monitored.

\section{CONCLUSIONS}

We have presented the first parsec-scale images of four candidate
southern GPS
sources and considered these alongside published VLBI images of the
five other candidate GPS sources from Edwards \& Tingay
\shortcite{edw04}.  We find core-jet or compact double morphologies
dominate, with further observations required to distinguish between
these two possibilities. The sole compact symmetric object, PKS
2254$-$367, is a nearby galaxy. One of the nine candidates, PKS
1831$-$711, displays appreciable variability, suggesting its GPS
spectrum is more ephemeral in nature. Although the sample is small,
there is some evidence that core-jet sources are more polarised and
more variable, and that compact doubles are the least variable.  MRC
1722$-$644, which has the narrowest spectral width of the sample, has
a compact double morphology, and while there are other examples of
such a correlation it is not universal.  Two of our sample have been
found to lie among the highest redshift GPS sources known.
An examination of the VLBI morphologies of the 15 catalogued
high-redshift ($z>$3) sub-class of GPS sources
suggests that core-jet morphologies predominate in this class,
with more systematic follow-up of this apparent trend encouraged.

\begin{acknowledgements}
The anonymous referee is thanked for suggestions which have
improved the paper.
The Long Baseline Array is part of the Australia Telescope National
Facility which is funded by the Commonwealth of Australia for
operation as a National Facility managed by CSIRO.  This research has
made use of NASA's Astrophysics Data System, and the NASA/IPAC
Extragalactic Database (NED) which is operated by the Jet Propulsion
Laboratory, California Institute of Technology, under contract with
the National Aeronautics and Space Administration.  We thank Dave
Jauncey, John Reynolds, Tasso Tzioumis and Jim Lovell in particular
for their efforts in operating and maintaining the LBA at the time
these observations were made.
\end{acknowledgements}

\end{document}